\documentclass[%
 reprint,
 amsmath,amssymb,
 aps,
showkeys]{revtex4-2}

\usepackage{graphicx}
\usepackage{dcolumn}
\usepackage{bm}
\usepackage{xcolor}
\usepackage{colortbl}

\begin{document}

\preprint{Physica A}





\title{Entropy of financial time series due to the shock of war}


\author{Ewa A. Drzazga-Szcz{\c{e}}{\'s}niak${^1}$}
\author{Piotr Szczepanik${^2}$}
\author{Adam Z. Kaczmarek${^3}$}%
\author{Dominik Szcz{\c{e}}{\'s}niak${^3}$}
\email{d.szczesniak@ujd.edu.pl}


\affiliation{
${^1}$Department of Physics, Faculty of Production Engineering and Materials Technology, Cz{\c{e}}stochowa University of Technology, 19 Armii Krajowej Ave., 42200 Cz{\c{e}}stochowa, Poland,\\
${^2}$Institute of Pricing and Market Analysis, Analitico, 49/8 Królewska Str., 47400 Racib{\'o}rz, Poland,\\
${^3}$Department of Theoretical Physics, Faculty of Science and Technology, Jan D{\l}ugosz University in Cz{\c{e}}stochowa, 13/15 Armii Krajowej Ave., 42200 Cz{\c{e}}stochowa, Poland}

\collaboration{Copernicus Institute Collaboration}

\date{\today}


\begin{abstract}

The concept of entropy is not uniquely relevant to the statistical mechanics but among others it can play pivotal role in the analysis of a time series, particularly the stock market data. In this area sudden events are especially interesting as they describe abrupt data changes which may have long-lasting effects. Here, we investigate the impact of such events on the entropy of financial time series. As a case study we assume data of polish stock market in the context of its main cumulative index. This index is discussed for the finite time periods before and after outbreak of the 2022 Russian invasion of Ukraine, acting as the sudden event. The analysis allows us to validate the entropy-based methodology in assessing market changes as driven by the extreme external factors. We show that qualitative features of market changes can be captured quantitatively in terms of the entropy. In addition to that, the magnitude of the impact is analysed over various time periods in terms of the introduced entropic index. To this end, the present work also attempts to answer whether or not the recent war can be considered as a reason or at least catalyst to the current economic crisis.

\end{abstract}

\keywords{entropy; volatility; information theory; econophysics; sudden events; war; time series; data science}

\maketitle


\section{Introduction}

In general, sudden or extreme events translate to the atypical patterns and deviations from the expected observations. As such, the ability to detect and address accordingly aforesaid anomalies is of great importance in various areas of science, technology or even social studies \cite{he2022, weinberg2017, aminikhanghahi2017, suriani2013, ramage1980}. This is to say, timing and occurrence of sudden events is essential when considering reliability of a system under extreme external conditions. A special attention to these aspects is given in the field of economy, where sudden events correspond to a notable incline/decline in economic activity or may even mark a breakdown of some economic models {\it e.g.} by exposing their limitations in terms of efficiency and rationality of the market \cite{evangelos2021}. In what follows, it is crucial to account for such events during economic modelling when considering processes such as the forecasting, decision-making or the anomaly detection. This is conventionally done on the grounds of the time series analysis, a vital part of data science \cite{montgomery2015}. The main reason for that is related to character of the time series itself, which are derived from the financial data and intrinsically encode information about economic events \cite{plerou2000}. Thus, to allow discussion of the extreme changes in economy, an appropriate tools in the time-series domain are required.

In the context of the above, entropy appears as an intriguing analytical concept, which spans beyond its original field of thermodynamics. While in terms of the statistical mechanics this property relates to the discrete probabilities of microstates, in the area of time series entropy is considered as an extension of the information theory \cite{rodriguez2022, sheraz2021, velichko2021, yin2016}, in accordance with the groundbreaking works by Shannon and Kolmogorov \cite{shannon1948, kolmogorov1998}. In particular, entropy can quantify the uncertainty, disorder or simply randomness of the time series, without adding constraints on the corresponding probability distribution \cite{yin2016, machado2010, shi2013, dionisio2006}. Hence, it constitutes an attractive alternative to the standard deviation for measuring market volatility \cite{bentes2008, bentes2012}, which is inherently sensitive to the sudden events or the {\it economic shocks}. In other words, since volatility relates to the degree of an asset movement over time, the sudden events are detectable in its spectrum. In this manner, entropy allows for discussing not only the magnitude of such fluctuations but also their distributions and patterns. It can account for the nonlinearities and correlations in the data sets, simultaneously capturing interdependence between assets \cite{darbellay2000, almog2019, lahmiri2020}. As a result, entropy constitutes potentially highly relevant framework for discussing impact of sudden events on the market and a pivotal tool in econophysics \cite{rodriguez2022, velichko2021, jakimowicz2020, yin2016, dionisio2006}.

So far, the studies on the economic sudden events in terms of entropy has been limited mainly to a few instances such as the investigations related to the 2008 economic crisis \cite{bose2012} or to the outbreak of the COVID-19 pandemic \cite{sheraz2021}. However, recent Russian invasion on Ukraine resulted in a yet another prominent economic shock, which is well-defined in terms of the time frames and influences multiple market branches. The economic consequences of this event constitute not only a perfect platform to investigate impact of the shock of war on the modern economy but also to validate the entropic methodology in assessing market changes due to the extreme external factors. Herein, we provide our contribution to this still not fully explored area. In details, we concentrate our study on the behavior of the main cumulative index of the polish stock market (WIG20) and conduct our calculations with respect to the conventional Shannon entropy. The WIG20 index is chosen due to the direct proximity of the corresponding market to the theater of war as well as the relatively high development of the polish economy. For convenience, the obtained results are compared with the predictions of the standard deviation. This analysis allows us to verify efficiency and predictive capabilities of the entropy-based formalism and to outline pertinent perspectives for the future research.

\section{Metodology}

The present analysis is conducted for the time-series of the daily log-returns, as calculated based on the financial data of interest. In particular, the daily log returns ($R_{i}$) are derived by following the relation:
\begin{equation}
R_i={\rm ln}\frac{P_{i}}{P_{i-1}}=\frac{P_{i}-P_{i-1}}{P_{i-1}}
\label{eq01},
\end{equation}
where $P_{i}$ ($P_{i-1}$) is the closing price of an asset on day $i$ ($i-1$). In this manner, we obtain convenient time-series data which is additive and symmetric in accordance to the scope of the present analysis.

The volatility of the above time-series is explored based on the two measures, namely: the standard deviation and the entropy. The former parameter is given by:
\begin{eqnarray}
S=\sqrt{\frac{1}{N-1}\sum_{i=1}^{N}(R_{i}-\mu)^2},
\label{eq02}
\end{eqnarray}
for the $N$ data points and $\mu$ being the arithmetic mean of all the returns. On the other hand the latter measure is calculated based on the Shannon entropy \cite{shannon1948}:
\begin{eqnarray}
H=-\sum_{i=1}^{M}p_{i}\ln{p_{i}}
\label{eq03}.
\end{eqnarray}
In Eq. (\ref{eq03}), $M$ stands for the number of bins (known also as the {\it intervals} or {\it classes} \cite{dogan2010}) in the discrete probability density function of the returns and $p_{i}$ is the probability corresponding to a given bin. Note that when in Eq. (\ref{eq03}) the logarithm base is $e$, the entropy is measured in {\it nats}. One can also use base equal to 2 or 10, resulting in the units of {\it shannons} or {\it hartleys}, respectively. In our case the change of units does not influences the qualitative behavior of entropy.

In the present study, the above theoretical model is feeded with the financial data of the WIG20 cumulative index and its composing stocks, as divided into two one-year-long data sets. The first set corresponds to the one-year time frame before the invasion (02/24/2021-02/23/2022), whereas the second considers similar period but after the beginning of the invasion (02/24/2022-02/23/2023). In what follows, we arrive with the total of $N=251$ data points for each set, providing sufficient economic perspective for our calculations. Note that the WIG20 index serves here as a pivotal parameter for comparison between the two approaches in modelling volatility. However, due to its cumulative character this index measures only the total fluctuations and to gain better insight into the underlying correlations of the market, the composing stocks are discussed. All of these stocks, included in the WIG20 index and the present analysis, are listed in Tab. \ref{tab01} along with their full names and market symbols. This list is valid for the assumed here time period but it is obviously a subject to changes in the future. For the sake of completeness, it is also crucial to remark that the component company Pepco was introduced to the stock market on 05/26/2021 {\it i.e.} the corresponding records does not cover the entire one-year period before the invasion. In addition, the composition of WIG20 index has changed four times over the analyzed time frame of two years. In details, on 03/18/2022 the already mentioned Pepco and other company named mBank replaced previously indexed stocks of Tauron and Mercator, respectively. Similarly, on 09/16/2022 company K{\c e}ty has replaced Lotos and on 12/16/2022 Kurk has switched with the PGING. All the described changes are appropriately marked in the results section.
 
To this end, for the purpose of the present study, the both data sets of interest are divided into the finite number of bins, which compose the discrete probability density function of the returns. There is no general and valid rule that determines the number and character of such bins \cite{dogan2010}. The final choice is always strongly related to the population of data points and their variability. In general, one should never stay with the empty bins or decrease their number to the point when resolution of the probability distribution is too low. In reference to the multiple models for the bin number, we observe that $M=20$ is optimal for our case and does not exceeds the upper theoretical limits for $N\sim250$ \cite{dogan2010}.

\section{Results}

\begin{figure*}
\includegraphics[width=\textwidth]{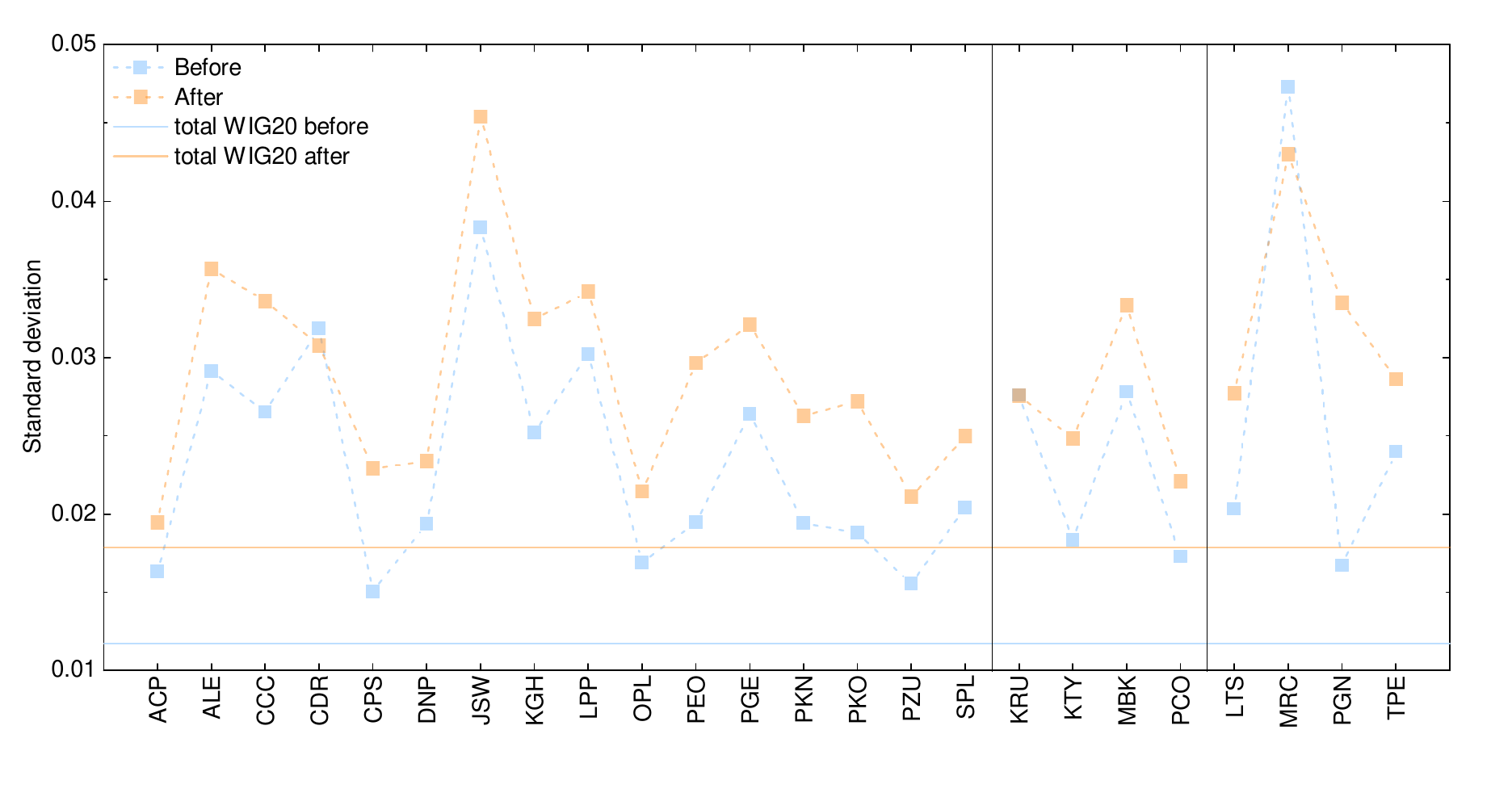}
\caption{\label{fig01} The standard deviation for the WIG20 index and its composing stocks. The first panel is for the constant component companies and the second (third) for the stocks introduced to (removed from) the index at some point. The results are given for the one-year time period before the beginning of the Russian invasion on Ukraine (blue) and after this event (orange). The solid lines correspond to the WIG20 index, whereas closed symbols represent estimates for the component stocks. Dashed lines are the guide for an eye.}
\end{figure*}

In Fig. \ref{fig01} we depict standard deviation as calculated for the WIG20 index and its composing stocks. According to the initial assumptions, the results are presented here for the one-year time frame before (orange color) and after (blue color) the beginning of the invasion. Note that Fig. \ref{fig01} is divided into three panels, first for the constant component companies, whereas second (third) panel is corresponding to the stocks which at some point were introduced to (removed form) the WIG20 index. For convenience the numerical results are additionally listed in Tab. \ref{tab01} together with the percentage difference between estimates obtained for the two considered time frames.

Upon the analysis of Fig. \ref{fig01}, the total standard deviation appears to be notably higher after the beginning of the invasion than for the time period before it. It means that the volatility of the market visibly increases for the former data set. In other words, this indicates high degree of stock price variations, which can be caused by not only the decline but also incline of the asset value. Similar behavior can be observed for most of the composing stocks. In details, one can notice that only companies like Kruk (debt management and purchase), Mercator (medical devices) and CD Projekt (video game developer and publisher) does not comply with this trend. The first company shows practically indistinguishable values for the two considered data sets, while the two latter ones present inverse behavior in comparison to the total standard deviation. The observed behavior for the first two companies is potentially related to the fact that their stocks where not included in the WIG20 index for the entire time, meaning their impact on the total index was limited. Moreover, Mercator capitalization, as a producer of medical gloves, was heavily reduced by the end of COVID-19 pandemic. Finally, the value of CD Projekt was a subject to turbulence due to the mixed reviews of their flagship video game product Cyberpunk 2077. Thus, the standard deviation for each of the three companies is the results of not only the war-time market changes but also other major factors.

\begin{figure}[hb!]
\includegraphics[width=\columnwidth]{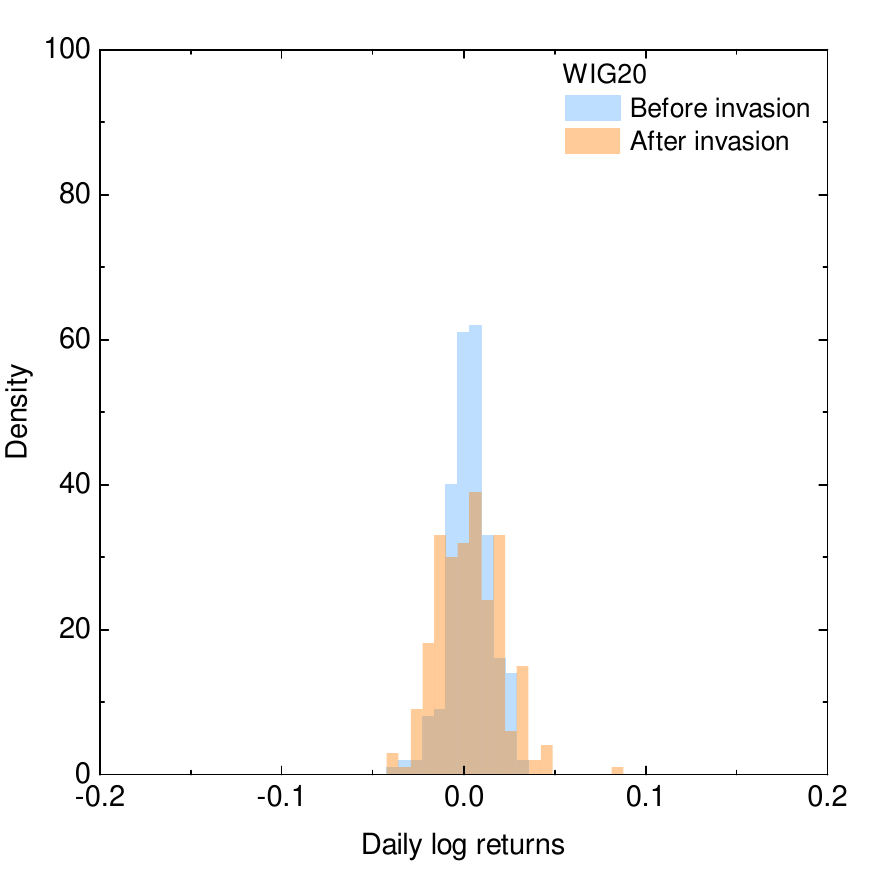}
\caption{\label{fig02} The discrete probability density function for the WIG20 index, the one-year period before (blue) and after (orange) the beginning of the Russian invasion on Ukraine.}
\end{figure}
\begin{figure*}
\includegraphics[width=\textwidth]{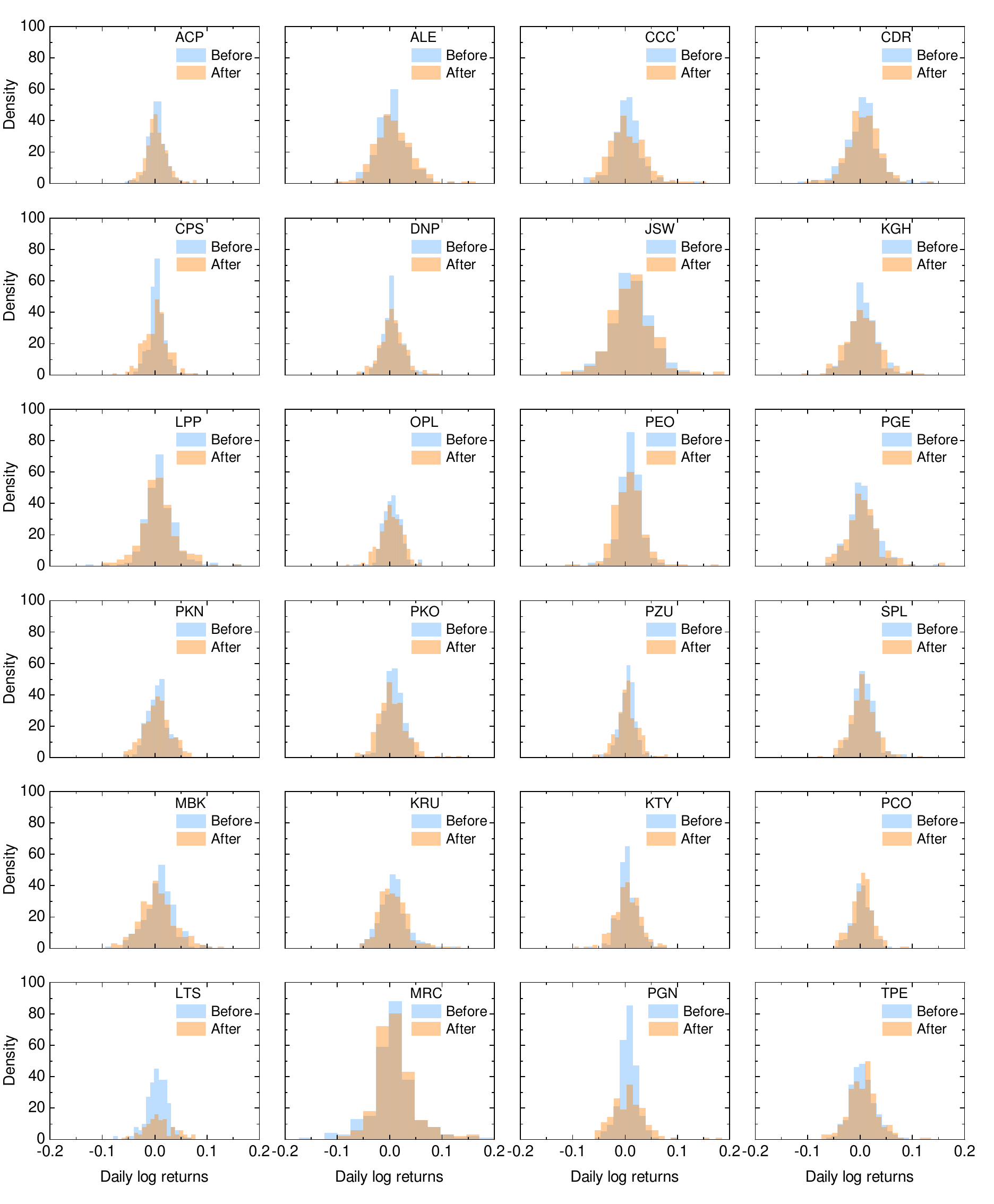}
\caption{\label{fig03} The discrete probability density function for the component stocks of the WIG20 index. The first four rows are for the constant component companies and the fifth (sixth) row is for the stocks introduced to (removed from) the index at some point. The results are presented for the one-year time period before the beginning of the Russian invasion on Ukraine (blue) and after this event (orange).}
\end{figure*}
\begin{figure*}[ht!]
\includegraphics[width=\textwidth]{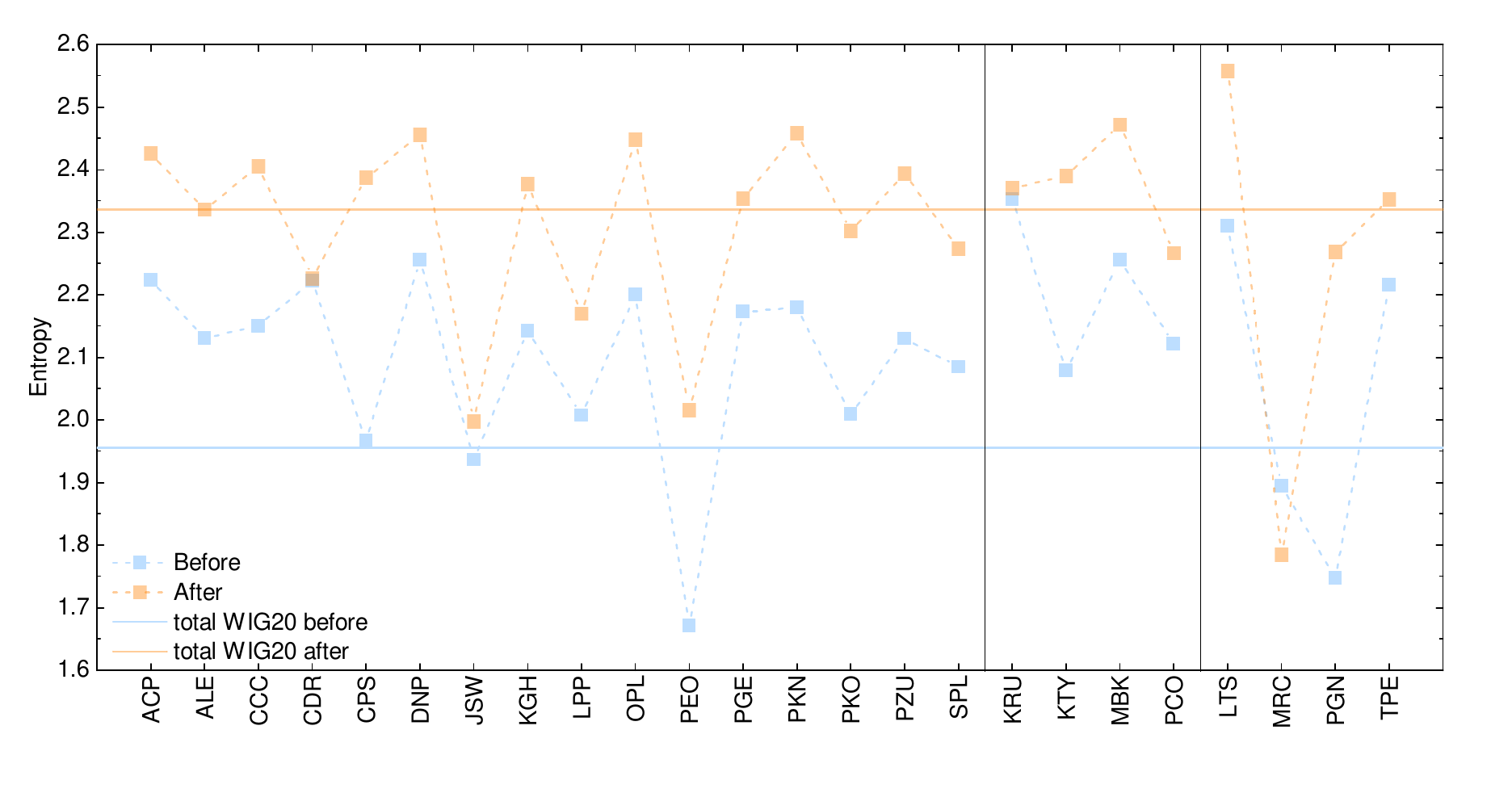}
\caption{\label{fig04} The Shannon entropy for the WIG20 index and its composing stocks. The first panel is for the constant component companies and the second (third) for the stocks introduced to (removed from) the index at some point. The results are given for the one-year time period before the beginning of the Russian invasion on Ukraine (blue) and after this event (orange). The solid lines correspond to the WIG20 index, whereas closed symbols represent estimates for the component stocks. Dashed lines are the guide for an eye.}
\end{figure*}

Nonetheless, the results for the individual stocks still allow us to observe that the largest volatility increase is present for the bank sector with other notable examples in petroleum and telecommunication sectors (see Tab. \ref{tab01} for details). Interestingly, by comparing the component estimates with the results for the cumulative index we can note that the total standard deviation measure does not average values obtained for the individual stocks. In fact, this measure is always lower that any of the corresponding component values. This is true for both considered sets of data and can originate from the way how the cumulative index is calculated or potentially from the shortcomings of the standard deviation approach. 

To investigate more in details the already observed trends, in Fig. \ref{fig02} and \ref{fig03} we present the discrete probability density function of the returns for the total index and its composing stocks, respectively. All the distributions are given for the time period before and after the beginning of the invasion, within the same color scheme as before. Based on Fig. \ref{fig02}, it can be observed that the probability distributions for the WIG20 index resemble normal distribution. However, the war-time data set is characterized by the much fatter tails and lower central maxima than the distribution corresponding to the index values before the conflict outbreak. This observation is in qualitative agreement with the results obtained within the standard deviation approach, which suggest higher volatility of the market after the beginning of the invasion. The situation is once again similar when inspecting return distributions for the component stocks {\it i.e.} volatility for most of the stocks is higher after the beginning of the Russian invasion. Still, there is some visible exception from this trend in terms of Pepco data. This is potentially due to the fact that, as mentioned earlier, data for Pepco does not cover the entire year before the invasion because of its relatively late introduction to the market on 05/26/2021.

\begin{figure*}[]
\includegraphics[width=\textwidth]{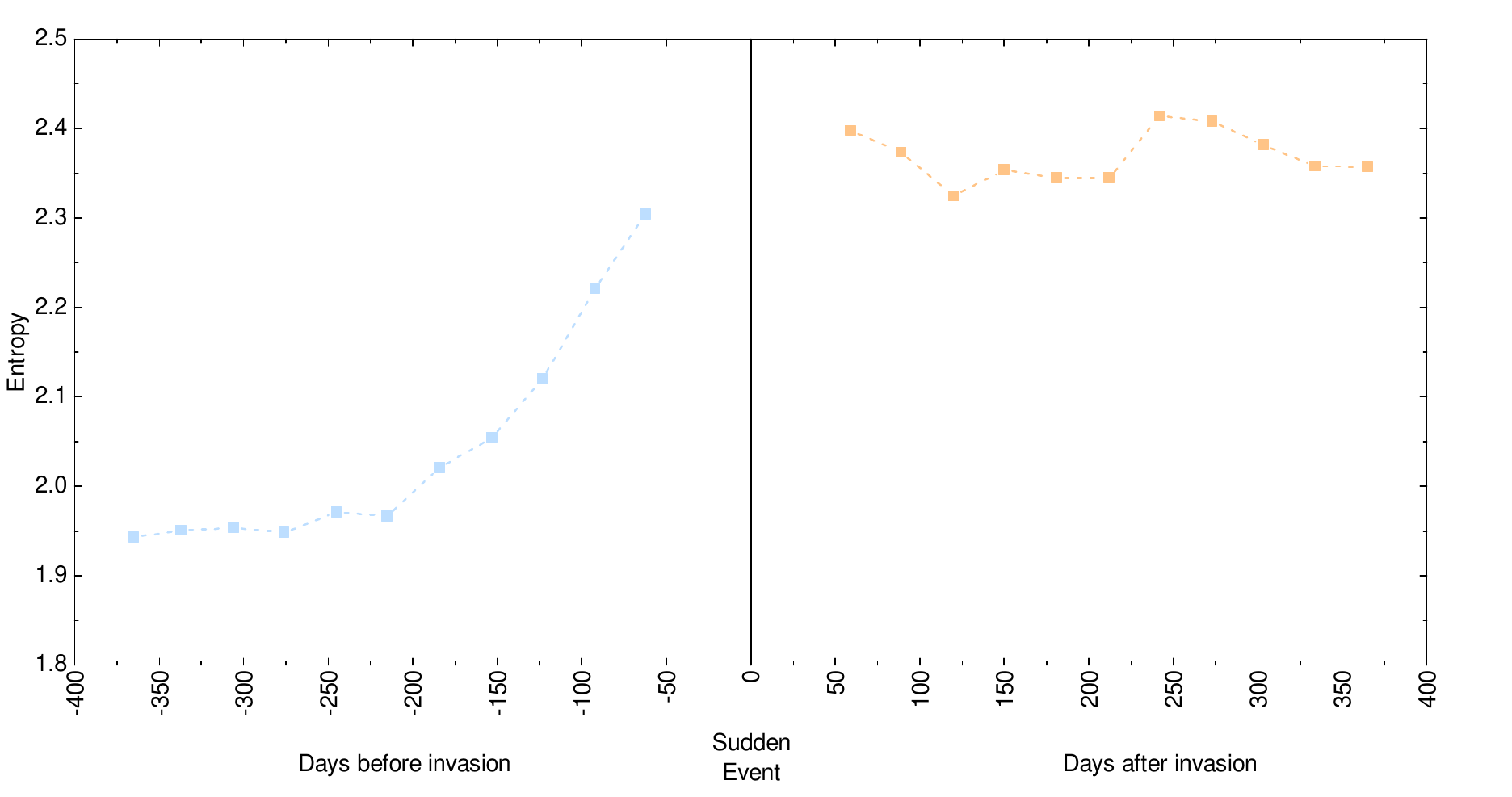}
\caption{\label{fig05} The Shannon entropy for the WIG20 index as calculated for different periods of time before (blue) and after (orange) the beginning of the Russian invasion on Ukraine. Dashed lines are the guide for an eye.}
\end{figure*}

It is next instructive to compare all the above results with the predictions of the entropic model. These are presented in Fig. \ref{fig04}, in a form of the entropy estimates for the WIG20 index and its component stocks, based on the two types of the data sets of interest. In general, the total entropy as well as the relative behavior between composing entropies is similar to the standard deviation predictions. However, closer inspection of the results allows to observe that, contrary to the previous case, here all the component stocks exhibit higher entropy after the war outbreak. The only exception is Mercator, relatively late in the WIG20 index and experiencing the COVID-19 related problems during the entire analyzed period as described before. Moreover, this time the results for the total index qualitatively averages results for the component companies. The mentioned observation is particularly visible for the data corresponding to the time frame after the beginning of the invasion. The results allows also to note that the percentage difference between results before and after invasion is smaller for each of the calculated entropies than in the case of the standard deviation results (see Tab. \ref{tab01}). Finally, the obtained entropies clearly follow character of the discrete probability density function in Fig. \ref{fig02} and \ref{fig03}.

To supplement our analysis we additionally plot the entropic index of WIG20 index for various time periods within the assumed here time frames. In Fig. \ref{fig05} we present the obtained results for the data sets before (left panel) and after (right panel) the beginning of the considered conflict. Both panels depict different behavior, namely: before the invasion the entropic index clearly increases when the assumed time distance from the invasion date becomes smaller, on the other hand the entropy is relatively stable throughout the entire period after the invasion data, independent of the number of considered days.

\begin{table*}
\caption{The numerical values of the standard deviation and entropy, as calculated for the WIG20 index and its composing stocks. The results are given for the one-year time period before the beginning of the Russian invasion on Ukraine and after this event. For convenience the percentage difference between estimates obtained for the two time frames of interest is given for the standard deviation and entropy. \label{tab01}}
\begin{ruledtabular}
\begin{tabular}{lccccccc}
Name & Symbol & Standard deviation & Standard deviation & Percentage & Entropy & Entropy & Percentage
\\  
& & before & after & difference & before & after & difference \\ \hline
Asseco & ACP & 0.016 & 0.019 & 17.632 $\%$ & 2.224 & 2.426 & 8.68 $\%$ \\
Allegro & ALE & 0.029 & 0.036 & 21.539 $\%$ & 2.131 & 2.336 & 9.178 $\%$
\\
CCC & CCC & 0.027 & 0.034 & 22.95 $\%$ & 2.151  & 2.405 & 11.15 $\%$
\\
CD Projekt & CDR & 0.032 & 0.031 & 3.175 $\%$ & 2.223 & 2.226 & 0.135 $\%$
\\
Cyfrowy Polsat & CPS & 0.015 & 0.023 &  42.105 $\%$ & 2.387 & 1.967 & 19.293 $\%$
\\
Dino & DNP & 0.019 & 0.023 & 19.048 $\%$ & 2.256 & 2.456 & 8.489 $\%$ \\
JSW & JSW & 0.038 & 0.045 & 16.868 $\%$ & 1.936 & 1.998 & 3.152 $\%$
\\
KGHM & KGH & 0.025 & 0.032 &  24.561 $\%$ & 2.143 & 2.378 & 10.39 $\%$
\\
LPP & LPP & 0.03 & 0.034 & 12.5 $\%$ & 2.008 & 2.17 & 7.755 $\%$
\\
Orange & OPL & 0.017 & 0.021 & 21.053 $\%$ & 2.201 & 2.449 & 10.66 $\%$ \\
Bank Pekao & PEO & 0.019 & 0.03 & 44.898 $\%$ & 1.671 & 2.016 & 18.714 $\%$ \\
PGE & PGE & 0.026 & 0.032 & 20.69 $\%$ & 2.173 & 2.353 & 7.954 $\%$
\\
Orlen & PKN & 0.019 & 0.026 & 31.11 $\%$ & 2.18 & 2.46 & 12.069 $\%$
\\
Bank PKO & PKO & 0.019 & 0.027 & 34.783 $\%$ & 2.01 & 2.302 & 13.544 $\%$
\\
PZU & PZU & 0.016 & 0.021 & 27.027 $\%$ & 2.13 & 2.393 & 11.629 $\%$
\\
Santander Bank & SPL & 0.02 & 0.025 & 22.222 $\%$ & 2.085 & 2.274 & 8.672 $\%$
\\
\hline
mBank & MBK & 0.028 & 0.033 & 16.393 $\%$ & 2.255 & 2.473 & 9.22 $\%$ \\
Kruk & KRU &  0.026 & 0.028 & 3.636 $\%$ & 2.353 & 2.371 & 0.762 $\%$
\\
K{\c e}ty & KTY & 0.018 & 0.025 & 32.558 $\%$ & 2.08 & 2.39 & 13.87 $\%$
\\
Pepco & PCO & 0.017 & 0.022 & 25.641 $\%$ & 2.121 & 2.267 & 6.655 $\%$ \\
\hline
LOTOS & LTS & 0.02  & 0.028  & 33.333  $\%$ &  2.31 & 2.558  & 10.189  $\%$
\\
Mercator & MRC & 0.047  & 0.043  & 8.889  $\%$ & 1.894  & 1.785  & 5.926  $\%$
\\
PGNiG & PGN & 0.017  & 0.033  & 64 $\%$ & 1.747  & 2.269  & 25.996 $\%$ \\
Tauron & TPE & 0.024  & 0.029  & 18.868  $\%$ & 2.216  & 2.352  & 5.954 $\%$ 
\end{tabular}
\end{ruledtabular}
\end{table*}

\section{Conclusions}

In the present study we have validated the entropy-based theoretical framework in describing behavior of financial time series under the influence of sudden and extreme external events. This has been done in the context of the WIG20 main cumulative index of the polish stock market for a one-year long data samples before and after the Russian invasion of Ukraine, respectively. In particular, it has been shown that entropy reproduces some of the features of the standard deviation, when describing the effects of the shock of war. In what follows, the obtained results confirmed that entropy can be indeed used as an alternative measure of volatility. These findings not only agree with the previous studies on applications of entropy in finances but also supplements them by considering the war-time driven changes in the stock market. For convenience all the numerical results are summarized in Tab. \ref{tab01}.

In addition to the above, the present study reveled several notable differences between entropy and standard deviation measures. First, the entropy was found to capture the character of empirical data, as depicted via the  discrete probability distribution function, in a more consistent way than the standard deviation. As a result, entropy was much better in highlighting differences between results obtained for the two time frames of interest. This was particularly visible in the case of CD Projekt data, where standard deviation predicted inverse behavior to the probability distribution function and entropy. Finally, it was also found that entropy of the cumulative index qualitatively average entropies of the composing stocks, again in contrast to the standard deviation estimates. This finding is particularly interesting since it shows that entropy holds potential in encompassing interdependecies between assets.

The last part of the analysis revealed that entropy measure can be used to quantify anomalies in time series toward their better detection. In particular, entropy exhibited different functional character when considering it for various time periods, before and after the beginning of the invasion. In other words, it can be argued that entropy showed signatures of the incoming economic shock. That means impact of potential sudden event can be visible in the entropy behavior when the time range is sufficiently small and context data is available for a long time range. In what follows, entropy may constitute building block for a future tools aimed at the sudden (extreme) event prediction. Interestingly, these results clearly showed also that the shock of war has a long lasting effect of an increased volatility of the market. At least in the one-year time perspective.

To this end, all the obtained results allow to make some preliminary judgements on the role of invasion in the current economic situation. Both the standard deviation and entropy points out that volatility of the polish market is higher after the crisis outbreak than before. In this manner, shock of war has visibly impacts polish economy. However, it is difficult to judge how big this impact is in comparison to other factors such as the still persisting effects of the COVID-19 pandemic and other internal economy-related decision of the government and related financial institutions. This would require an additional investigations.

\appendix

\bibliography{apssamp}

\end{document}